\documentclass[%
superscriptaddress,
reprint,
preprintnumbers,
amsmath,amssymb,aps,prl]{revtex4-1}

\usepackage{bm}
\usepackage[colorlinks=true,linktocpage=true,linkcolor=blue,citecolor=blue,urlcolor=blue]{hyperref}

\usepackage{cleveref} 

\usepackage[utf8]{inputenc} 
\usepackage[T1]{fontenc} 
\usepackage{microtype}

\usepackage[mathscr]{euscript}

\usepackage{mathtools}

\usepackage{tensor}  

\usepackage{color,soul}

\newcommand{\half}{\frac{1}{2}}
\newcommand{\nn}{\nonumber}

\newcommand{\dbrk}{\right. \\ \left.}

\newcommand{\df}{\mathrm{d}}
\newcommand{\dow}{\partial}

\usepackage{xparse}

\ExplSyntaxOn

\clist_map_inline:nn{A,B,C,D,E,F,G,H,I,J,K,L,M,N,O,P,Q,R,S,T,U,V,W,X,Y,Z}
{
  \expandafter\def\csname b#1\endcsname{\bm{#1}} 
  
  \expandafter\def\csname c#1\endcsname{\mathcal{#1}} 
  
  \expandafter\def\csname bc#1\endcsname{\bm{\mathcal{#1}}} 
  
  \expandafter\def\csname s#1\endcsname{{\mathsmaller{#1}}} 
  
  \expandafter\def\csname bb#1\endcsname{\mathbb{#1}} 
  
  \expandafter\def\csname rm#1\endcsname{\mathrm{#1}} 
  
  \expandafter\def\csname sc#1\endcsname{\mathscr{#1}} 
  
  \expandafter\def\csname sf#1\endcsname{\mathsf{#1}} 
  
  \expandafter\def\csname f#1\endcsname{\mathfrak{#1}} 
}

\ExplSyntaxOff

\newcommand{\lB}{\left [}
\newcommand{\rB}{\right ]}
\newcommand{\lb}{\left (}
\newcommand{\rb}{\right )}

\newcommand\ext{\text{ext}}

\usepackage{blindtext}
\usepackage{mathbbol}

\DeclareRobustCommand{\DIEP}{\ensuremath{%
\mathchoice{\includegraphics[height=2ex]{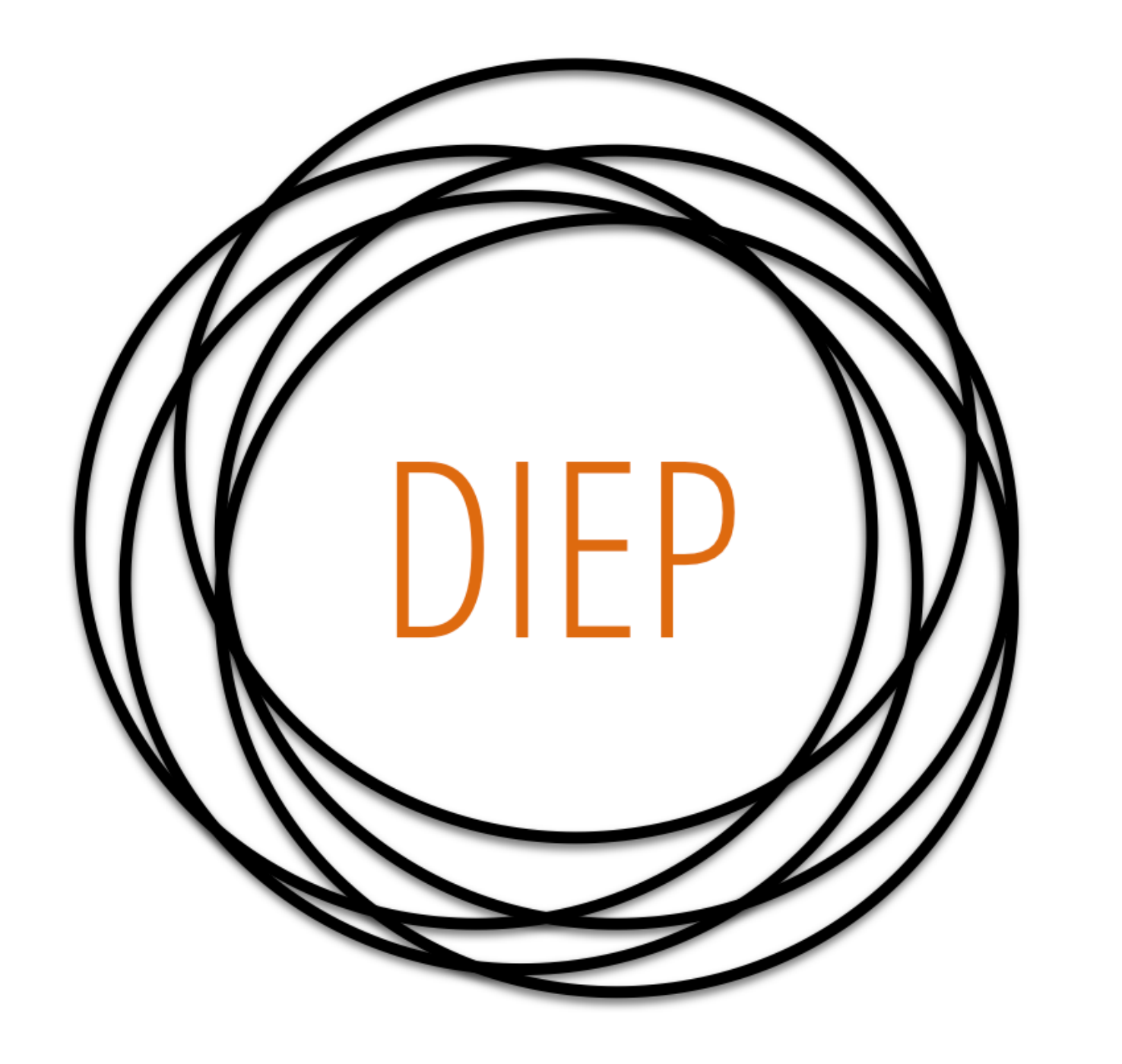}}
  {\includegraphics[height=2ex]{DIEPs.pdf}}
    {\includegraphics[height=1.5ex]{DIEPs.pdf}}
    {\includegraphics[height=1ex]{DIEPs.pdf}}
  }}

\begin{document} 

\title{Hydrodynamics for charge density waves and their holographic duals}

\author{Jay Armas} 
\email{j.armas@uva.nl}
\affiliation{Institute for Theoretical Physics, University of Amsterdam,
  1090 GL Amsterdam, The Netherlands}
\affiliation{\DIEP Dutch Institute for Emergent Phenomena, The Netherlands}

\author{Akash Jain}
\email{ajain@uvic.ca}
\affiliation{Department of Physics \& Astronomy, University of Victoria, PO Box 1700 STN CSC, Victoria, BC, V8W 2Y2, Canada.}


\begin{abstract}
  We formulate a theory of dissipative hydrodynamics with spontaneously broken translations, describing charge density waves in a clean isotropic electronic crystal. We identify a novel linear transport coefficient, lattice pressure, capturing the effects of background strain and thermal expansion in a crystal. We argue that lattice pressure is a generic feature of systems with spontaneously broken translations and must be accounted for while building and interpreting holographic models. We also provide the first calculation of the coefficients of thermal and chemical expansion in a holographic electronic crystal. 
\end{abstract}

\pacs{Valid PACS appear here}

\maketitle


Ever since the discovery of high-temperature superconductivity, cuprates continue to be enigmatic owing to their intricate phase diagrams exhibiting various intertwined patterns of symmetry breaking \cite{Fradkin_2015, NatureZaanen}. In particular, the phase diagram of copper oxides contains electronic liquid crystal phases that spontaneously break translations and/or rotations. These include the elastic multi-component charge density wave (CDW) phases \footnote{This phase is also known as Wigner crystal phase (see e.g. \cite{Delacretaz:2017zxd} ).}, smectic stripe phases, where the CDW pattern only appears along a single direction, or nematic spin density wave phases, where translations are intact but rotations are spontaneously broken. All these phases can potentially appear simultaneously with superconducting phases where the global U(1) symmetry is also spontaneously broken (see~\cite{Gruner:1988zz, Fradkin_2015, NatureZaanen} for a comprehensive review). To note is the fact that CDW ground states are an essential feature across the phase diagram of copper oxides.

Electrons in cuprates, in particular in strange metallic phases, are strongly-correlated. This renders the quasi-particle Fermi liquid crystal theory unreliable for these systems, even qualitatively, leaving us with only a handful of techniques for this plethora of phases~\cite{NatureZaanen}. Recently, hydrodynamics has been proposed as a theoretical framework for studying aspects of strongly correlated electron systems \cite{Delacretaz:2016ivq, Delacretaz:2017zxd, Delacretaz:2019wzh}, capable of explaining pinning in the optical conductivity and predicting the magnitude of viscosity in optimally doped BSCCO \cite{Delacretaz:2017zxd}. Another series of efforts has been directed towards holography~\cite{Amoretti:2017axe, Amoretti:2017frz, Gouteraux:2018wfe, Amoretti:2018tzw, Amoretti:2019kuf, Donos:2018kkm, Donos:2019tmo, Donos:2019hpp}, where properties of strongly coupled quantum systems are being probed using classical gravity. In fact, within this setting, hydrodynamics is directly related to such holographic models via the fluid/gravity correspondence~\cite{Bhattacharyya:2008jc}.

However, all previous treatments of hydrodynamics for charged lattices (see e.g.~\cite{chaikin_lubensky_1995, PhysRevA.6.2401, PhysRevB.22.2514, Delacretaz:2017zxd}) have not considered an essential transport coefficient in their constitutive relations, namely the \emph{lattice pressure}. This coefficient first appeared in \cite{Armas:2019sbe} in the context of uncharged viscoelastic materials, and models a uniform repulsion/attraction between lattice sites in a material with translational order. However, the thermodynamic variation of lattice pressure can be understood as carrying information about the thermal expansion of the lattice: \emph{coefficients of thermal} and \emph{chemical expansion} \footnote{Earlier works (\cite{chaikin_lubensky_1995, PhysRevA.6.2401, PhysRevB.22.2514, Delacretaz:2017zxd}) did not include lattice pressure, which allows the description of thermodynamically disfavorable states, however, they include certain susceptibilities that capture the effects of expansion coefficients at zero lattice pressure. In particular, \cite{Delacretaz:2017zxd} set these coefficients to zero in their analysis. Nevertheless, the effect of expansion coefficients on correlation functions or mode propagation in the presence of lattice pressure has not been explicitly derived before.}. As also discussed in \cite{Armas:2019sbe}, lattice pressure is generically present in holographic models of viscoelasticity. 

The main purpose of this letter is to provide the complete hydrodynamic theory for isotropic charged crystals, including contributions from lattice pressure. We derive the hydrodynamic predictions for linear modes and response functions. As far as we are aware, the sound and diffusion modes in the longitudinal sector for charged crystals have not been previously worked out in full generality in the literature. We also comment on the signatures of lattice pressure in holography using a simple class of holographic models. Our analysis illustrates that many previous works have used an incomplete hydrodynamic framework to interpret holographic results in CDW (e.g. \cite{Amoretti:2017axe, Amoretti:2017frz, Gouteraux:2018wfe, Amoretti:2018tzw, Amoretti:2019kuf}), as in  viscoelasticity (e.g. \cite{Andrade:2019zey, Baggioli:2019mck, Baggioli:2019abx, Ammon:2019apj, Baggioli:2018bfa})\footnote{It was also noted in \cite{Donos:2019hpp} that this framework was also incomplete for describing strong explicit translation symmetry breaking. This, however, is unrelated to the framework we consider here. }. We derive an analytic formula for the coefficients of thermal and chemical expansion in these simple models.

For clarity of presentation and to effectively focus on the impact of lattice pressure, we restrict our attention to clean CDW phases. That is, we do not consider the effects of pinning or momentum dissipation due to interactions with the ionic lattice, or the presence of topological defects such as disclinations and dislocations.


\vspace{1em}

\emph{Crystal field theory \& lattice pressure.}---The fundamental ingredient in an effective theory for crystals is a set of crystal fields $\phi^I$. They represent the spatial distribution of lattice cores within the crystal~\cite{Armas:2019sbe} and can be understood as Goldstones of spontaneously broken translations. The indices $I,J,\ldots = 1,\ldots, k \leq d$ run over the number of broken translations, while $\mu,\nu,\ldots = 0,\ldots d$ run over spacetime indices. Physical distances between the cores are measured by $h^{IJ} = g^{\mu\nu}e^I_\mu e^J_\nu$, where $e^I_\mu = \dow_\mu\phi^I$ and $g_{\mu\nu}$ is the background metric. The $I,J,\ldots$  indices are  raised/lowered using $h^{IJ}$ and $h_{IJ} = (h^{-1})_{IJ}$. The crystal also carries a ``preferred'' reference configuration $\mathbb{h}_{IJ} = \delta_{IJ}/\alpha^2$ where $\alpha$ is a constant parametrising the ``inverse size'' of the crystal. Distortions of the crystal away from this reference configuration are measured by the non-linear strain tensor $u_{IJ} = (h_{IJ} - \mathbb{h}_{IJ})/2$.

The free energy of a crystal in an isotropic phase, up to quadratic order in small strain expansion, takes the form $F = - \int \df^d x\sqrt{-g}\, P$ with
\begin{multline}
    \label{eq:Pressure}
    P
    = P_f 
    + P_\ell\lb u^I{}_{\!I} + u^{IJ}u_{IJ} \rb \\
    - \half B\,  (u^I{}_{\!I})^2 
    - G \lb u^{IJ}u_{IJ} - \frac{1}{d} (u^I{}_{\!I})^2 \rb
    + \mathcal{O}(u^3).
\end{multline}
Here $P_f$ is the thermodynamic or ``fluid'' pressure and $P_\ell$ is the lattice pressure, while $B$ and $G$ are bulk and shear modulus respectively. Classical elasticity theory usually describes thermodynamically stable states, requiring the free energy to be minimised with respect to strain and setting the linear term $P_\ell|_{\text{eq}}=0$ in equilibrium~\cite{landau1986theory}. However, in the context of various holographic models, one finds that $P_\ell|_{\text{eq}}\ne0$. As argued in \cite{Amoretti:2017axe}, such states can be relevant for strange metallic regions where quantum critical fluctuations of the order parameters do not provide any stable ordered phase. Furthermore, even in states with $P_\ell|_{\text{eq}}=0$, thermodynamic derivatives of $P_\ell$ are generically nonzero and measure the coefficients of thermal and chemical expansion (see \S6 of~\cite{landau1986theory})
\begin{equation}
\label{eq:expansion_coeffs}
    \alpha_{T} = \frac{1}{B} \frac{\dow P_\ell}{\dow T},\qquad 
    \alpha_\mu = \frac{1}{B} \frac{\dow P_\ell}{\dow \mu}.
\end{equation}
These derivatives are shown to leave nontrivial signatures in the hydrodynamic spectrum (see e.g.~\cite{Armas:2019sbe, Ammon:2020xyv}).


\vspace{1em}

\emph{Viscoelastic hydrodynamics.}---We are interested in low-energy fluctuations of a charged crystal around thermal equilibrium. In addition to $\phi^I$, the dynamics in this regime is governed by conserved operators: stress tensor $T^{\mu\nu}$ and charge/particle current
$J^\mu$
\begin{equation} \label{eq:elastic}
    \nabla_\mu T^{\mu\nu}
    = F^{\nu\rho} J_\rho - K^\ext_I e^{I\nu}\,, \quad
    \nabla_\mu J^\mu = 0\,.
  \end{equation}
Here $A_\mu$ and $K_I^\ext$ are background sources coupled to $J^\mu$ and $\phi^I$, while $g_{\mu\nu}$ is the source for $T^{\mu\nu}$. $F_{\mu\nu} = 2\dow_{[\mu}A_{\nu]}$. Collectively, these determine time-evolution of the hydrodynamic fields: velocity $u^\mu$ (with $u^\mu u_\mu = -1$), temperature $T$, and chemical potential $\mu$. The most generic set of constitutive relations for $T^{\mu\nu}$ and $J^\mu$ for an isotropic~\footnote{In the anisotropic case there are other transport coefficients which are either odd-rank or antisymmetric rank-2 in $I,J,\ldots$ indices.} charged viscoelastic fluid at one-derivative order in Landau frame are given as
\begin{align} \label{eq:const_elastic}
  J^\mu
  &= q u^\mu 
  - P^{I\mu} \sigma^q_{IJ} P^{J\nu} \lb
    T\dow_\nu \frac{\mu}{T} - E_\nu \rb 
    - P^{I\mu} \gamma_{IJ} u^\nu e^J_\nu\,,\nn\\
  T^{\mu\nu}
  &= (\epsilon + P) u^\mu u^\nu + P g^{\mu\nu}
    - r_{IJ} e^{I\mu} e^{J\nu}  \nn\\
    &\qquad 
    - P^{I(\mu} P^{J\nu)} \eta_{IJKL}  P^{K(\rho} P^{L\sigma)}
    \nabla_{\rho}u_\sigma\,.
\end{align}
Here, $P$ is the thermodynamic pressure, $\epsilon$, $q$, and $s$ are the energy, charge, and entropy densities, and $r_{IJ}$ is the elastic stress tensor. All these quantities are functions of $T$, $\mu$, and $h^{IJ}$. They obey the thermodynamic relations: $\df P = s \df T + q \df \mu + \half r_{IJ} \df h^{IJ}$ and $\epsilon + P = sT + q\mu$. We have defined $P^{\mu\nu} = g^{\mu\nu} + u^\mu u^\nu$, $P^{I\mu}=P^{\mu\nu}e^I_\nu$,  $E_\mu = F_{\mu\nu}u^\nu$. Furthermore, $\eta_{IJKL}$, $\sigma^q_{IK}$, and $\gamma_{IK}$ are dissipative transport coefficient matrices. In addition, the constitutive relations have to be supplemented with configuration equations determining the time-evolution of $\phi^I$, i.e.
\begin{multline}\label{eq:phiI-EOM}
    \sigma^\phi_{IJ} u^\mu \dow_\mu \phi^I
    + \gamma'_{JK} P^{K\mu} 
    \lb T\dow_\mu \frac{\mu}{T} - F_{\mu\nu} u^\nu \rb\\
    + \nabla_\mu \lb r_{JK} e^{K\mu} \rb
    = K_J^\ext\,.
\end{multline}
Here, $\sigma^\phi_{IK}$ and $\gamma'_{IK}$ are two more matrices of dissipative transport coefficients. At zeroth order in derivatives, these equations imply that the crystal fields are constant along the fluid flow. Taking $\phi^I = \alpha(x^I - \delta\phi^I)$, they turns into their more familiar form
$u^t \dow_t\delta\phi^I = u^I - u^i\dow_i\delta\phi^I + \ldots$.

Following our discussion in~\cite{Armas:2019sbe}, it can be checked that \cref{eq:const_elastic,eq:phiI-EOM} above are the most generic set of constitutive relations and configuration equations that satisfy the local second law of thermodynamics, $\nabla_\mu S^\mu \geq 0$, with the entropy current $S^\mu = s\, u^\mu - \frac{\mu}{T}(J^\mu - q u^\mu)$, provided that the symmetric parts of
\begin{equation}
    \label{eq:2ndLawIneqGeneral}
    \eta_{(IJ),(KL)}\,, \quad 
    \begin{pmatrix}
    \sigma^q_{IK} & \gamma_{IK} \\
    \gamma'_{IK} & \sigma^\phi_{IK}
    \end{pmatrix}\,,
\end{equation}
are positive semidefinite matrices.



\vspace{1em}

\emph{Linear regime.}---We are typically interested in crystals close to mechanical equilibrium, where we can expand the hydrodynamic equations in small strain. The pressure $P$ can be expanded as in \cref{eq:Pressure}, which determines $q$, $s$, $\epsilon$, and $r_{IJ}$ up to linear order in strain through thermodynamics. At one derivative order, we only keep the strain-independent terms, i.e. 
\begin{gather}
    \sigma^q_{IJ}=\sigma_q h_{IJ},~
    \sigma^\phi_{IJ}=\sigma_\phi h_{IJ},~
    \gamma_{IJ}=\gamma\, h_{IJ},~
    \gamma'_{IJ}=\gamma' h_{IJ}, \nn\\ 
    \eta_{IJKL}= \left(\zeta-{\textstyle\frac{2}{d}}\eta\right)h_{IJ}h_{KL}
    + 2\eta\, h_{IK}h_{JL}\,.
\end{gather}
We can identify $\eta$ and $\zeta$ as shear and bulk viscosities, $\sigma_q$ as charge conductivity, $\sigma_\phi$ as crystal diffusivity, while $\gamma$, $\gamma$' as certain mixed conductivities. The second law constraints in \cref{eq:2ndLawIneqGeneral} reduce to
\begin{equation}\label{eq:linear-inequalities}
  \eta, \zeta, \sigma_q, \sigma_\phi \geq 0~~, \qquad
  \sigma_q\sigma_\phi \geq \frac14 \lb\gamma+\gamma'\rb^2.
\end{equation}
Finally, we arrive at the constitutive relations in the small-strain regime
\begin{align}\label{eq:linear-isotropic-consti}
  J^\mu
  &= \lb q_f + q_\ell u^\lambda{}_{\!\!\lambda}\rb\! u^\mu
    - \sigma_q P^{\mu\nu} \!\lb
    T \dow_\nu \frac{\mu}{T} - E_\nu \rb
    - \gamma P^{\mu}_I u^\nu e^I_\nu, \nn\\
  T^{\mu\nu}
  &= \lb \epsilon_f
  + \epsilon_\ell u^\lambda{}_{\!\!\lambda} \rb u^\mu u^\nu
    + \lb P_f + P_\ell u^\lambda{}_{\!\!\lambda}\rb P^{\mu\nu}
    + P_\ell h^{\mu\nu} \nn\\
    - \eta\,&\sigma^{\mu\nu}
    - \zeta P^{\mu\nu} \dow_\rho u^\rho 
    - 2 G u^{\mu\nu}
    - \lb B- {\textstyle\frac{2}{d}}G\rb u^\lambda{}_{\!\!\lambda} h^{\mu\nu}.
\end{align}
Here $h_{\mu\nu} = h_{IJ} e^I_\mu e^{J}_\nu$ and $u_{\mu\nu} = u_{IJ} e^I_\mu e^{J}_\nu$. Similarly the configuration equations \eqref{eq:phiI-EOM} reduce to
\begin{align}\label{eq:linear-phiI-EOM}
    &\sigma_\phi u^\mu e^I_\mu 
    - h^{IJ}\nabla_\mu \lb P_\ell e^{\mu}_J
    - \lb B - {\textstyle\frac{2}{d}}G\rb u^\lambda{}_{\!\!\lambda} e^{\mu}_J
    - 2 G u^{\mu\nu} e_{J\nu} \rb \nn\\
    &\qquad 
    + \gamma' P^{I\mu} \lb
    T\dow_\mu \frac{\mu}{T} - E_\nu u^\nu \rb
    = h^{IJ} K_J^\ext\,.
\end{align}
We have defined the fluid thermodynamics $\df P_f = s_f \df T + q_f \df \mu$, $\epsilon_f + P_f = s_fT + q_f\mu$ and similarly for the lattice pressure $\df P_\ell = s_\ell \df T + q_\ell \df \mu$, $\epsilon_\ell + P_\ell = s_\ell T + q_\ell \mu$. Setting $u_{IJ}=0$, note that the mechanical pressure $\langle T^{xx}\rangle = P_f+P_\ell$ gets contribution from both thermodynamic and lattice pressure.


\vspace{1em}

\emph{Conformality.}---Let us briefly comment on the conformal limit of our theory, due to its relevance in holography. Requiring that the stress tensor scales appropriately leads to the conformality constraints at the non-linear level: $\epsilon = d\, P - r_{IJ}h^{IJ}$ and $h^{IJ}\eta_{IJKL}=\eta_{IJKL}h^{KL} = 0$. In the linear regime, they imply 
\begin{equation}\label{eq:conformal-linear-constraints}
  \epsilon_f = d\lb P_f + P_\ell \rb,\quad
  \epsilon_\ell = d\lb P_\ell - B \rb, \quad 
  \zeta = 0.
\end{equation}
Notice that having $P_\ell$ or $\epsilon_\ell$ non-zero in the theory (unlike~\cite{Delacretaz:2017zxd}), allows for a non-zero $B$ in a conformal crystal. Furthermore, using the expansion coefficients from \cref{eq:expansion_coeffs}, we can derive the identity
\begin{equation}
    T\alpha_T + \mu \alpha_\mu = (d+1)\frac{P_\ell}{B} - d.
    \label{eq:expansion_identity}
\end{equation}
In particular, in a state with no lattice pressure or chemical potential,  $\alpha_T < 0$. This is not surprising, as the size of a conformal crystal scales inversely with temperature at constant $\mu/T$.


\vspace{1em}

\emph{Linear hydrodynamics and modes.}---Consider a charged crystal on flat spacetime, $g_{\mu\nu} = \eta_{\mu\nu}$, with trivial external sources, $A_\mu = \mu_0\delta^t_\mu$, $K^\ext_I = 0$. An equilibrium configuration on this background is given by $T=T_0$, $\mu=\mu_0$, $u^\mu = \delta^\mu_t$, $\phi^I = \alpha\, x^I$. We can expand \cref{eq:elastic,eq:phiI-EOM} linearly in fields around this configuration to obtain the constitutive relations of linear hydrodynamics. We recover the previously known results of~\cite{Delacretaz:2017zxd, Amoretti:2019cef} with the identification $\xi = 1/\sigma_\phi$, $\gamma_1 = -\gamma/\sigma_\phi$, and $\sigma_0 = \sigma_q + \gamma^2/\sigma_\phi$, only if we choose $\gamma'=-\gamma$ and set lattice pressure $P_\ell$ and both its derivatives $s_\ell$, $q_\ell$ to zero \footnote{Effects of $s_\ell$, $q_\ell$ in a state with $P_\ell|_{eq} = 0$ can be captured by certain couplings mentioned in appendix A of~\cite{Delacretaz:2017zxd}, which have been switched off by the authors for simplicity.}.

Solving the linear equations in momentum space, we can find the complete set of linear modes admitted by the theory. We find two pairs of sound modes, one each in transverse and longitudinal sectors, and two diffusive modes in the longitudinal sector
\begin{equation}
  \omega = \pm v_{\parallel,\perp} k 
  - \frac{i}{2} \Gamma_{\parallel,\perp} k^2 + \ldots\,, \quad
  \omega = -i D^{q,\phi}_\parallel k^2 + \ldots.
\end{equation}
In the transverse sector, one finds that the modes take a simple form known previously~(eg. \cite{Delacretaz:2017zxd})
\begin{equation}
  v_\perp^2 = \frac{G}{\chi_{\pi\pi}}\,, \qquad
  \Gamma_\perp
  =  \frac{w_f^2}{\chi_{\pi\pi}^2} \frac{G}{\sigma}
  + \frac{\eta}{\chi_{\pi\pi}}\,,
\end{equation}
where $\chi_{\pi\pi} = \epsilon_f + P_f + P_\ell$ is the momentum susceptibility and $w_f = \epsilon_f + P_f$ is the enthalpy density. The transverse speed $v_\perp$ is controlled by the shear modulus $G$; in the $G=0$ case this mode reduces to the well known shear diffusion mode in hydrodynamics. Modes in the longitudinal sector are considerably more involved. With applications to holography in mind, we present the results for conformal viscoelastic fluids here for simplicity. The general non-conformal results are given in the supplementary material. The longitudinal sound mode simplifies in this limit to
\begin{gather}
  v_\parallel^2
  = \frac1d
  + \frac{2 \frac{d-1}{d} G}{\chi_{\pi\pi}}\,,~~
  \Gamma_\parallel
  =  \frac{w_f^2   \lb 2 \frac{d-1}{d} G \rb^2}{\sigma^\phi \chi^3_{\pi\pi} v_\parallel^2}
  + \frac{2\frac{d-1}{d} \eta}{\chi_{\pi\pi}}\,.
\end{gather}
This is the usual sound mode present in hydrodynamics, but gets modified on a lattice. Longitudinal diffusion modes, on the other hand, are given by the solutions of the quadratic  
\begin{multline}
  \lb D_\parallel
  - \frac{w_f^2}{\sigma_\phi}
  \frac{2\frac{d-1}{d}G + B - P_\ell}
  {d\,\chi_{\pi\pi} v_\parallel^2(w_f+w_\ell)} \rb
  \lb \frac{\Xi\, D_\parallel}{d(w_f+w_\ell)} - \frac{\sigma_q}{T^2} \rb \\
  =\frac{D_\parallel}{\sigma_\phi}
  \lb \frac{s_f q_\ell - q_f s_\ell}{w_f+w_\ell} + \frac{\gamma}{T} \rb
  \lb \frac{s_f q_\ell - q_f s_\ell}{w_f+w_\ell} - \frac{\gamma'}{T} \rb,
\end{multline}
where $\Xi = \frac{\dow s_f}{\dow T}\frac{\dow q_f}{\dow \mu}
- \frac{\dow s_f}{\dow \mu}\frac{\dow q_f}{\dow T}$ and $w_\ell = \epsilon_\ell + P_\ell$. The two modes are controlled by the coefficients $\sigma_q$, $\sigma_\phi$: in the $\sigma_\phi\to\infty$ limit we recover the usual charge diffusion mode $D_\parallel^q$, but modified on a lattice, while in the $\sigma_q\to0$ limit we obtain the uncharged crystal diffusion mode $D_\parallel^\phi$ characteristic of a lattice \footnote{$D^\phi_\parallel$ is the standard crystal diffusion, related to the fact that mass motion can be different from lattice motion \cite{PhysRevA.6.2401}. In our hydrodynamic formulation, it bears no relation to quasicrystals as suggested in \cite{Baggioli:2020nay}.} (see~\cite{PhysRevA.6.2401}).

We note that, in the conformal case, $P_\ell$ appears explicitly only in the diffusion modes (modulo the implicit dependence in $\chi_{\pi\pi} = \langle T^{tt}\rangle + \langle T^{xx}\rangle$). Therefore, if we were to ignore $P_\ell$, for instance as in~\cite{Delacretaz:2017zxd}, hydrodynamics would lead to incorrect predictions for diffusion modes (see~\cite{Ammon:2020xyv} for a particular example in holographic massive gravity). For non-conformal theories, however, $P_\ell$ infects all the modes in the longitudinal sector explicitly.


\vspace{1em}

\emph{Response functions and Onsager's relations.}---We can compute retarded two-point functions in our model by solving the hydrodynamic equations \eqref{eq:linear-isotropic-consti} and \eqref{eq:linear-phiI-EOM} in presence of infinitesimal plain wave sources~\cite{Kovtun:2019hdm}. Working at zero wave vector, we find in the full non-conformal case
\begin{gather} 
  G^R_{T^{xx}T^{xx}}
  = \chi_{\pi\pi}v_\parallel^2
    - i\omega \lb \zeta + 2\frac{d-1}{d}\eta \rb
    + \langle T^{xx} \rangle, \nn\\
  G^R_{T^{xy}T^{xy}}
  = G
    - i\omega\eta
    + \langle T^{xx} \rangle \,, \nn\\
  G^R_{J^x J^x}
  = \frac{q_f^2}{\chi_{\pi\pi}}
  - i \omega \tilde\sigma_q, \quad 
  G^R_{\phi^x\phi^x}
  = \frac{1}{\omega^2\chi_{\pi\pi}}
    + \frac{\tilde\sigma_\phi}{i\omega} \nn\\
  G^R_{J^x \phi^x}
  = - \frac{q_f}{i\omega\chi_{\pi\pi}} + \tilde\gamma , \quad 
  G^R_{\phi^x J^x}
  = \frac{q_f}{i\omega\chi_{\pi\pi}}
  + \tilde\gamma',
\end{gather}
where we have defined the dissipative response coefficients 
\begin{gather}
    \tilde\sigma_q
    = \sigma_q 
    + \frac{1}{\sigma_\phi} 
    \lb \frac{q_f P_\ell}{\chi_{\pi\pi}} - \gamma\rb
    \lb \frac{q_f P_\ell}{\chi_{\pi\pi}} + \gamma'\rb,~~
    \tilde\sigma_\phi = \frac{w_f^2}{\sigma_\phi\chi_{\pi\pi}^2}, \nn\\
    \tilde\gamma 
    = \frac{w_f}{\sigma_\phi}
    \lb \frac{\gamma}{\chi_{\pi\pi}} - \frac{q_f P_\ell}{\chi_{\pi\pi}^2}\rb,~~
    \tilde\gamma' = \frac{w_f}{\sigma_\phi}
    \lb \frac{\gamma'}{\chi_{\pi\pi}} + \frac{q_f P_\ell}{\chi_{\pi\pi}^2}\rb \nn.
\end{gather}
All the remaining response functions are either zero or related to these by isotropy. For $P_\ell = 0$ and $\gamma = -\gamma'$, these results reduce to the expressions reported in~\cite{Amoretti:2019cef}, up to contact terms.

If the system enjoys $\Theta=\text{T}$ (time-reversal) or $\Theta=\text{PT}$ (spacetime parity) invariance, Onsager's relations require $G^R_{J^x \phi^x} = - \Theta G^{R}_{\phi^x J^x}$, setting $\gamma = -\gamma'$. This is the case assumed in~\cite{Amoretti:2019cef}. In the case of $\Theta=\text{CPT}$ invariance, however, $G^R_{J^x \phi^x} = \Theta G^R_{\phi^x J^x}$ and we instead have $\gamma'|_{\mu\to-\mu} = \gamma$ (note that $q_f$ flips sign under CPT).


\vspace{1em}

\emph{Holography.}---As an application of our hydrodynamic theory, we propose a simple holographic model for clean CDW phases following the discussion in~\cite{Armas:2019sbe,Baggioli:2014roa}. We also compute the coefficients of thermal and chemical expansion in this model. Specialising to four bulk dimensions, the model is described by Einstein-Maxwell gravity in the bulk coupled to two scalars
\begin{align} \label{eq:action}
  S_{\text{bulk}}
  &= \frac12 \int \df^{4} x \sqrt{-G}\lb R + 6
     -\frac{1}{4}\mathcal{F}^2 - 2 V(X) \rb.
\end{align}
Here $G_{ab}$ is the bulk metric with $a,b,...$ being the bulk indices, and $\mathcal{F}_{ab}=2\partial_{[a}\mathcal{A}_{b]}$ is the field strength associated with the gauge field $\mathcal{A}_a$. Here $V(X) = X + \ldots$ is an arbitrary potential in $X=\delta_{IJ} \half G^{ab}\dow_a \Phi^I \dow_b \Phi^J$ for a set of scalar fields $\Phi^I$. To describe a thermal state at the boundary, we consider charged black brane solutions of the action \eqref{eq:action} of the form
\begin{gather}
  \df s^2
  = \frac{1}{r^2f(r)} \df r^2 + r^2
  \lb - f(r) \df t^2 + \delta_{IJ} \df x^I \df x^J \rb, \nn\\
  \mathcal{A}_\mu =\mu \left(1-\frac{r_0}{r}\right)\delta^t_\mu, \qquad
  \Phi^I = \alpha\, x^I,
  \label{eq:sol}
\end{gather}
where $r_0$ is the horizon radius, $r\to\infty$ is the conformal boundary, $\mu$ the chemical potential, and $\alpha$ an arbitrary constant. The blackening factor is given by
\begin{equation}
 f(r)= 1 - \frac{r_0^{3}}{r^{3}}
 -\frac{r_0\mu^2(r-r_0)}{4r^4} - \frac{\alpha^2}{r^{3}}
  \int_{r_0}^r \df r' \frac{V(X(r'))}{X(r')},
\end{equation}
where $X(r)=\alpha^2/r^2$. The profile of the scalars breaks the translational invariance in the boundary theory. This model, with $V(X)=X$, has been considered in the context of momentum dissipation in~\cite{Andrade:2013gsa} with explicitly broken translations. However, contrary to \cite{Andrade:2013gsa}, we introduce alternative boundary conditions for $\Phi^I$ so as to describe spontaneously broken translations at the boundary. We will also allow for an arbitrary renormalisation scale parameter $\cM$ in the boundary conditions breaking the conformal symmetry of the model. The holographic renormalisation procedure along with the choice of boundary counter-terms is detailed in the supplementary material.

Identifying the onshell action as free-energy for the boundary theory, we can read out the thermodynamic pressure in equilibrium 
\begin{equation}
    P_f = \frac{r_0^3}{2} \Big(
    1 + \frac{\mu^2}{4r_0^2} + 2U_0 - V_0 \Big)
    - \alpha^2 \cM.
\end{equation}
Here $U(X)=\frac{-1}{2}X^{3/2}\int dX X^{-5/2}V(X)$ and $X_0 = X(r_0)$, along with $V_0= V(X_0)$, $V'_0 = V'(X_0)$, and $U_0 = U(X_0)$.  We can also extract the stress tensor, charge current, and scalar expectation values using the boundary behaviour of the solution and read out
\begin{gather}
  \epsilon_f = 
   r_0^3 \lb 1+\frac{\mu^2}{4r_0^2} - U_0 \rb
   + \alpha^2\cM, \quad 
   q_f = \frac{\mu r_0}{2},
   \nn \\
   s_f = 2\pi r_0^2, \quad
   P_\ell = \frac{r_0^3}{2} \lb V_0 - 3U_0\rb + \alpha^2\cM,
   \label{eq:hol_thermo}
\end{gather}
along with $\phi^I=\alpha\, x^I$. With temperature defined as $T = r_0^2 f'(r_0)/(4\pi)$, one can check that the expected thermodynamic relations given below \cref{eq:linear-phiI-EOM} are satisfied. We can easily obtain the bulk modulus by deforming our solution from $\alpha \to \alpha+\delta\alpha$, leading to a uniform strain $u_{IJ} = -\alpha \delta_{IJ} \delta\alpha$, and using \cref{eq:Pressure}. We find
\begin{equation}
    B = \frac{3r_0^3}{4} \lb V_0 - 3U_0\rb
    + \frac{\frac12 X_0 V'_0 r^3_0 \lb 3 - V_0 + \frac{\mu^2}{4r_0^2} \rb}
    {3-V_0+2X_0V'_0+\frac{\mu^2}{4r_0^2}}
    + \alpha^2\cM.
\end{equation}
One can check that these expressions satisfy the conformal identities \eqref{eq:conformal-linear-constraints} in the absence of $\cM$, confirming that $\cM$ characterises RG flow away from the conformal fixed point. Expressions for $G$ and the dissipative transport coefficients have to be obtained numerically.

Finally, we can use the expressions for $B$ and $P_\ell$ to read out the expansion coefficients in \cref{eq:expansion_coeffs}. In particular, in a conformal model with $\cM=0$ around a state with zero lattice pressure $P_\ell|_{eq} = 0$, we simply find 
\begin{equation}
    \alpha_T = 
    \frac{-4s_f}
    {2s_f T + q_f\mu}, \qquad 
    \alpha_\mu =
    \frac{-2q_f}{2s_f T + q_f\mu},
\end{equation}
irrespective of the model dependent potential $V(X)$. They follow the conformal identity \eqref{eq:expansion_identity}. We find both the expansion coefficients to be negative for our holographic conformal crystal. This behaviour is altered for $\cM\neq 0$. However, negative thermal expansion is not unusual in solid materials \cite{doi:10.1088/1468-6996/13/1/013001}.


A lattice configuration is thermodynamically stable if it minimises the free energy: $(\delta P_f/\delta \alpha)_{T,\mu} = -2/\alpha\, P_\ell|_{eq} = 0$ for some $\alpha\neq 0$. Equivalently, $\langle T^{xx} \rangle$ must be equated to $P_f$~\cite{Donos:2013cka}. Notice that $P_\ell\ne0$ in a generic equilibrium configuration in \eqref{eq:hol_thermo}. This is also the case for similar holographic models with spontaneously broken translations~\cite{Andrade:2019zey, Baggioli:2019mck, Baggioli:2019abx, Ammon:2019apj}. In fact, simple monomial models with $V(X) = X^N$ and $\cM = 0$, do not admit any thermodynamically stable configurations. Fortunately, one can consider polynomial models, such as $V(X)=X+\lambda X^2$ or the ``higher-derivative model'' of~\cite{Amoretti:2017frz}, that do admit thermodynamically stable configurations, in our case $\alpha^2=r_0(r_0-\cM)/(2\lambda)$. Though the lattice pressure $P_\ell$ is zero in such configurations, its thermodynamic derivatives $s_\ell$, $q_\ell$, $\epsilon_\ell$ are still generically nonzero and have to be taken into account in the hydrodynamic spectrum. This was verified for the uncharged case in~\cite{Ammon:2020xyv}. Previous holographic models for CDW have not been taking the lattice pressure into account, leading to the misinterpretation of some of their results.


\vspace{1em}

\emph{Outlook.}---We have provided a complete formulation of hydrodynamics for clean isotropic CDW phases, taking into account the new transport coefficient $P_\ell$. We find that $P_\ell$ non-trivially modifies the longitudinal sector of linear fluctuations. Besides being crucial for correctly interpreting the holographic results, including those of \cite{Amoretti:2017axe, Amoretti:2017frz, Gouteraux:2018wfe, Amoretti:2018tzw, Amoretti:2019kuf}, lattice pressure is also highly relevant for real condensed matter systems. It can describe parts of the phase diagram for which there are no thermodynamically stable ordered phases and also accounts for the effects of thermal expansion of the crystal. We have obtained an analytic expression for the coefficients of thermal and chemical expansion is a class of simple holographic models using lattice pressure.

It will be interesting to further include the effects of explicit translation symmetry breaking (momentum dissipation and pinning) as well as incorporate spontaneous breaking of U(1) global symmetry. This would provide a more robust theory for realistic scenarios. In this context, it would be relevant to revisit some of the results and predictions of \cite{Delacretaz:2016ivq, Delacretaz:2017zxd, Delacretaz:2019wzh, Amoretti:2019buu} with our understanding of lattice pressure, potentially including weak/strong background magnetic fields. In particular, it is an open question whether the existing data can constrain the magnitude of $P_\ell$ or its gradients for specific materials. It would also be interesting to work out an analogous formulation for smectic and nematic charged liquid crystal phases.

In the context of holography, we focused on equilibrium thermal states dual to planar black hole geometries. However, this work provides the necessary linear transport theory for interpreting near-equilibrium states. By computing the quasinormal modes and using the Kubo formulae reported here, one can extract all first order transport coefficients and check whether the modes reported here reproduce holographic results. We leave some of these explorations for future work.


\vspace{1em}

We would like to thank M. S. Golden, B. Gout\'eraux and P. Kovtun for various helpful discussions and comments. JA is partly supported by the Netherlands Organization for Scientific Research (NWO). AJ is supported by the NSERC Discovery Grant program of Canada.

\bibliography{mySpiresCollaboration_JAAJ,chargedvisco}


\appendix
\renewcommand\thefigure{A.\arabic{figure}}    
\setcounter{figure}{0} 
\renewcommand{\theequation}{A.\arabic{equation}}
\setcounter{equation}{0}

\vspace{1em}

\section{\large Supplementary Material}

In this appendix we provide details on 
the complete set of longitudinal dispersion relations for non-conformal charged isotropic crystals, as well as the holographic renormalisation procedure employed in the letter. 
\vspace{1em}


\emph{Longitudinal modes.}---In the main text, we presented longitudinal modes in case of a conformal charged viscoelastic fluid for simplicity. Here we provide the generic non-conformal expressions. Let us first change the thermodynamic variables from $(T,\mu)$ to $(\epsilon_f,q_f)$ via
\begin{gather}
    \frac{\dow T}{\dow \epsilon_f} 
    = \frac{1}{T\Xi} \frac{\dow q_f}{\dow \mu}, \qquad 
    \frac{\dow T}{\dow q_f} 
    = -\frac{1}{T\Xi} \frac{\dow \epsilon_f}{\dow \mu}, \nn\\
    \frac{\dow \mu}{\dow \epsilon_f} 
    = -\frac{1}{T\Xi} \frac{\dow q_f}{\dow T}, \qquad 
    \frac{\dow \mu}{\dow q_f} 
    = \frac{1}{T\Xi} \frac{\dow \epsilon_f}{\dow T},
\end{gather}
where $\Xi
  = \frac{\dow s_f}{\dow T}\frac{\dow q_f}{\dow \mu}
  - \frac{\dow s_f}{\dow \mu} \frac{\dow q_f}{\dow T}$. 
The longitudinal sound velocity is given as
\begin{equation}
  v_\parallel^2
  = \frac{\lb w_f + w_\ell \rb \frac{\dow P_m}{\dow \epsilon_f}
  + \lb q_f + q_\ell \rb \frac{\dow P_m}{\dow q_f}}{\chi_{\pi\pi}}
  + \frac{B + 2 \frac{d-1}{d}G - P_\ell}{\chi_{\pi\pi}},
\end{equation}
whereas the attenuation is
\begin{align}
  \Gamma_\parallel
  &=  
  \frac{
    \lb \frac{\dow P_m}{\dow q_f}\rb^2  \sigma_q + \frac{1}{\sigma_\phi} 
    \lb \mathcal{F}_1 + \frac{\dow P_m}{\dow q_f} \gamma \rb
  \lb \mathcal{F}_1 - \frac{\dow P_m}{\dow q_f} \gamma' \rb }{v_\parallel^2\chi_{\pi\pi}} \nn\\
  &\qquad 
  + \frac{\zeta + 2\frac{d-1}{d}\eta}{\chi_{\pi\pi}}.
\end{align}
Here $P_m = P_f + P_\ell$ is the mechanical pressure. 
For clarity of presentation, we have defined 
\begin{gather}
  \mathcal{F}_1
  = w_f \lb v_\parallel^2 - \frac{\dow P_m}{\dow \epsilon_f} \rb
    - q_f \frac{\dow P_m}{\dow q_f}, \quad 
  \mathcal{F}_2
  = T^2\frac{\dow(\mu/T)}{\dow q_f}.
\end{gather}
The quadratic governing the diffusion modes, on the other hand, is given by
\begin{widetext}
\begin{multline}
  \frac{D_\parallel}{T\cF_2} \lb D_\parallel 
  - \frac{w_f^2}{\sigma_\phi}
  \frac{2\frac{d-1}{d}G + B - P_\ell}
  {\chi_{\pi\pi} v_\parallel^2 \Xi T\mathcal{F}_2} \rb
  - \frac{\sigma_q}{T^2}
  \lB \lb \frac{(w_f+w_\ell)^2}{\chi_{\pi\pi} \Xi T\mathcal{F}_2}
  + \frac{2\frac{d-1}{d}G + B - P_\ell}{\chi_{\pi\pi}}
   \rb \frac{ D_\parallel}{v_\parallel^2}
  - \frac{w_f^2}{\sigma_\phi}
  \frac{2\frac{d-1}{d}G + B - P_\ell}
  {\chi_{\pi\pi} v_\parallel^2 \Xi T\mathcal{F}_2}
  \rB
  \\
  = \frac{D_\parallel}{\sigma_\phi v_\parallel^2}
  \lB 
  \frac{(w_f+w_\ell)^2}{\chi_{\pi\pi}T \Xi \cF_2}
  \lb \frac{s_f q_\ell - q_f s_\ell}{w_f+w_\ell} + \frac{\gamma}{T} \rb
  \lb \frac{s_f q_\ell - q_f s_\ell}{w_f+w_\ell} - \frac{\gamma'}{T} \rb
  \dbrk
  + \frac{B + 2\frac{d-1}{d}G - P_\ell}{\chi_{\pi\pi} } 
  \lb \frac{\dow P_f/\dow q_f}{\cF_2} - \frac{\gamma}{T} \rb
  \lb \frac{\dow P_f/\dow q_f}{\cF_2} + \frac{\gamma'}{T} \rb
  \rB.
\end{multline}
\end{widetext}


\vspace{1em}

\emph{Holographic renormalisation.}---In order for the action \eqref{eq:action} to be finite on the class of black brane solutions in \eqref{eq:sol}, we need an additional Gibbons-Hawking-York counter-term at the boundary along with a boundary potential for the scalars. To wit,
\begin{equation}
  S_{\text{counter}}
  = \int_{r=r_c} \df^3 x \sqrt{-\gamma}
    \lb K - 2 + \bar V(\bar X) \rb,
\end{equation}
where we have defined the induced metric $\gamma_{\mu\nu}$ at the location of the cutoff surface $r=r_c$. We have assumed the boundary to be flat and avoided any curvature dependent terms. Here $\bar X = \delta_{IJ} \half \gamma^{\mu\nu}\dow_\mu \Phi^I \dow_\nu \Phi^J$ and $K = G^{ab}\mathrm{D}_a n_b$ is the mean extrinsic curvature, where $n_a$ is an outward pointing normal vector to the surface and $\mathrm{D}_a$ the covariant derivative compatible with the bulk metric $G_{ab}$. For the onshell action not to have any divergences, the potential must take the form
\begin{equation} \label{eq:vpot}
  \bar V(\bar X)
  = 2 \lb 1 - \sqrt{1 - U(\bar X)} \rb
  - \sum_n \frac{\mathcal{M}_n}{r_c^3} (r_c^2 \bar X)^n.
\end{equation}
In \eqref{eq:vpot} we have introduced ${\mathcal{M}}_n$, which are arbitrary cutoff dependent renormalisation scale parameters taken to be regular in the limit $r_c\to\infty$. Their presence spoils the conformal symmetry of the holographic model by imposing non-conformal boundary conditions. Different values of ${\mathcal{M}}_n$ describe different physical theories, or different points in the RG flow of the same physical theory. In the letter, we have  only turned on $\cM = {\mathcal{M}}_1$ for simplicity.

Assuming the bulk potential falling as $V(X) \sim X \sim r^{-2}$ near the boundary, we can read out the boundary value of the fields
\begin{equation}
    g_{\mu\nu} = \lim_{r_c\to\infty} \frac{1}{r_c^2} \gamma_{\mu\nu},\quad
    A_\mu=\lim_{r_c\to\infty} \mathcal{A}_\mu, \quad 
    \phi^I = \lim_{r_c\to\infty} \Phi^I.
    \label{eq:hol_fields}
\end{equation}
For potentials falling off more quickly, the qualitative behaviour of the scalars is different; see~\cite{Alberte:2017oqx} for more details. Fields in \cref{eq:hol_fields} serve as sources for the respective operators obtained by varying the total onshell action
\begin{align}
  &\delta S^{\text{onshell}}_{\text{bulk+counter}} \nn\\
  &= \int_{~~\mathclap{~r=r_c}} \df^{3} x \sqrt{-g}\lb
    \half T^{\mu\nu} \delta g_{\mu\nu}+J^{\mu}\delta A_\mu
    - \Pi_I \delta \phi^I \rb,
\end{align}
where
\begin{align}
  T^{\mu\nu}
  &= \lim_{r_c\to\infty} r^{5}_c
  \bigg( K \gamma^{\mu\nu} - K^{\mu\nu} - 2 \gamma^{\mu\nu} \nn\\
  &\qquad 
    + \bar V(\bar X) \gamma^{\mu\nu}
    - \delta_{IJ} \bar V'(\bar X)
    \gamma^{\mu\rho} \gamma^{\nu\sigma} \dow_\rho \Phi^I \dow_\sigma \Phi^J 
    \bigg), \nn\\
   J^\mu
   &=\lim_{r_c\to\infty}\frac{r_c^{3}}{2}\mathcal{F}^{u a}n_a, \nn\\
  \Pi_I
  &= \lim_{r_c\to\infty} r_c^{3} \delta_{IJ}
    \bigg( V'(X) n^a\dow_a \Phi^J \nn\\
    &\qquad 
    + \frac{1}{\sqrt{-\gamma}}
    \dow_\mu \lb \sqrt{-\gamma}\, \gamma^{\mu\nu} \bar V'(\bar X) \dow_\nu \Phi^J \rb \bigg).
\end{align}
We have arrived at the holographic formula for $T^{\mu\nu}$ and $J^\mu$. In this picture, however, the fields $\phi^I$ serve as sources leading to explicit breaking of translations. In order to describe spontaneous symmetry breaking, we deform the theory with a surface action $S_{\text{alternative}} = \int \df^{3}x \sqrt{-g}\, \Pi_I \phi^I$ implementing alternative quantisation and switching the roles of $\Pi_I$ and $\phi^I$. In this picture, $K^\ext_I \equiv \Pi_I$ are the background sources coupled to the dynamical fields $\phi^I$. Finally, the boundary conditions imposed on our holographic model for spontaneously broken translations are $g_{\mu\nu} = \eta_{\mu\nu}$, $A_\mu = \mu \delta^t_\mu$, and $\Pi_I = 0$. 

\clearpage

\end{document}